\documentstyle[aps,prb,epsf,epsfig,twocolumn]{revtex}

\begin{document}

\draft

\twocolumn[\hsize\textwidth\columnwidth\hsize\csname @twocolumnfalse\endcsname

\title{Charge-density waves in the Hubbard chain: evidence for 4$k_F$ instability}
\author{Thereza Paiva$^{\, (1)}\!$ and Raimundo R.\ dos Santos$^{\, (2)}$}
\address{$^{(1)}$Department of Physics, University of California, Davis, California, 
         95616-8677\\         
         $^{(2)}$Instituto de F\' \i sica, Universidade Federal do Rio de Janeiro,
                 Cx.P.\ 68.528, 21945-970 Rio de Janeiro RJ, Brazil\\}
                 
\date{\today}

\maketitle

\begin{abstract}
Charge density waves in the Hubbard chain  are studied
by means of finite-temperature Quantum Monte Carlo simulations and Lanczos
diagonalizations for the ground state.
We present results both for the charge susceptibilities and for the charge
structure factor at densities $\rho=1/6$ and 1/3; 
for $\rho=1/2$ (quarter filled)
we only present results for the charge structure factor. 
The data are consistent with a $4k_F$ instability dominating over the 
$2k_F$ one, at least for sufficiently large values of the Coulomb 
repulsion, $U$.
This can only be reconciled with the Luttinger liquid analyses if the
amplitude of the $2k_F$ contribution vanishes above some $U^*(\rho)$.
\end{abstract}

\pacs{PACS:     
      71.27.+a, 
      71.10.-w, 
      71.45.Lr, 
      72.15.Nj, 
      73.20.Mf. 
}
\vskip2pc]

Charge-density waves (CDW's) are present in a variety of strongly 
correlated electron systems, ranging from quasi-one-dimensional organic 
conductors\cite{JS82,Pouget89,Jerome94} to the more recently discovered 
manganites.\cite{Cheong}
In order to understand the influence of CDW formation on magnetic order and 
transport properties, a crucial issue is to establish the 
period of charge modulation in the ground state. 
This period, in turn, is primarily determined by the interplay between 
electron-phonon and electron-electron couplings. 
For the specific example of quasi-one dimensional organic conductors at 
quarter-filled band, both period-2 and period-4 modulations have 
been observed;\cite{Pouget89}
these correspond, respectively, to $4k_F$ and $2k_F$, where 
$k_F=\pi\rho/2$ is the Fermi wave vector for a density $\rho$ of free 
electrons on a periodic lattice.
The use of simplified effective models capturing the basic physical ingredients 
should therefore be extremely helpful in predicting the dominant 
instability.\cite{Ung94}
In this context, the Hubbard model can be thought of as a limiting case 
(of vanishing electron-phonon interaction), in which the influence of 
electronic correlations on CDW modulation can be monitored.
However, even for this simplest possible model there has been a disagreement
between analyses of the continuum (Luttinger liquid) 
version,\cite{Solyom79,Schulz90,Frahm90,Voit94} and
early finite temperature (world-line) quantum Monte Carlo (QMC) 
simulations.\cite{HS}
According to the former, the large-distance behaviour of the charge
density 
correlation function is given by\cite{Schulz90}
\begin{equation}
\langle n(x)n(0)\rangle = {K_\rho\over (\pi x)^2} 
           + A_1 { \cos (2 k_F x) \over x^{1+K_\rho} \ln^{3/2} x}\\
           + A_2 { \cos (4 k_F x) \over x^{4K_\rho}},
\label{nn}
\end{equation}
where the amplitudes $A_1$ and $A_2$, and the exponent $K_\rho$ are 
interaction- and density-dependent parameters; 
for repulsive interactions\cite{Schulz90} ${1\over 2} \leq K_\rho < 1$, 
so that charge correlations are expected to be dominated by the $2k_F$ term.
By contrast, the simulations pointed towards $4k_F$ being the
main correlations. 
Nonetheless, based on a Renormalization Group (RG) analysis, it was 
argued\cite{HS} that the $2k_F$ instability should
eventually dominate over the $4k_F$ one for sufficiently low temperatures.
For infinite coupling the system becomes effectively a spinless fermion
problem with a $4k_F$ instability,\cite{Schulz90,HS,Ogata90} and there
is no disagreement in this case.

Since present day computational capabilities allow one to reach much lower 
temperatures and larger system sizes than before, a numerical reanalysis of 
the model is certainly in order. 
Our purpose here is to present the results of such a study.
The Hubbard Hamiltonian reads
\begin{equation}
{\cal H}=-t\sum_{i,\sigma} 
 c_{i\sigma}^{\dagger}c_{i+1\sigma}^{\phantom{\dagger}}+
         U\sum_i n_{i\uparrow}^{\phantom{\dagger}} 
         n_{i\downarrow}^{\phantom{\dagger}},
\label{c-ham}
\end{equation}
where, in standard notation, $U$ is the on-site Coulomb repulsion;
the hopping integral sets the energy scale, so we take $t=1$ throughout this
paper. 
We probe finite temperature properties through
determinantal QMC simulations\cite{bss,Hirsch85,lg92,vdl92} for the 
grand-canonical version of (\ref{c-ham}), 
$\hat{\cal H}\equiv {\cal H} -\mu \hat{N}$, 
where 
$\hat{N}\equiv \sum_{i\sigma} n_{i\sigma}$;
the chemical potential $\mu$ is adjusted to yield the desired particle density.
This analysis is supplemented by zero temperature calculations:
the ground state of Eq.\ (\ref{c-ham}), for finite 
lattices of $N_s$ sites with periodic boundary conditions,
is obtained through the Lanczos algorithm,\cite{Roomany80,Gagliano86,Dagotto94}
in the subspace of fixed particle-density (canonical ensemble). 

The signature of a CDW instability is a {\em peak} at $q=q^*$ in the 
zero-temperature limit of the charge-density susceptibility,
\begin{equation}
N(q)={1\over N_s}\int_0^\beta d\tau \sum_{i,\ell} 
     \langle\;  n_i(\tau) n_{i+\ell}(0)\;  \rangle\; e^{iq\ell},
\label{Nq}
\end{equation} 
where the imaginary-time dependence of the operators is given by 
$n_i(\tau)\equiv e^{\tau\hat{\cal H}}\; n_i\; e^{-\tau\hat{\cal H}}$,
with $n_i = n_{i\uparrow} + n_{i\downarrow}$.
We recall that in simulations `time' is discretized in intervals 
$\Delta\tau$, such that 
the size along this direction is $L=\beta/\Delta\tau$.
The CDW instability should also show up as a {\em cusp}, again at $q=q^*,$ in the 
zero-temperature charge-density structure factor,
\begin{equation}
C(q)={1\over N_s}\sum_{i,\ell}
\langle 0|\; n_i n_{i+\ell}\; |0\rangle\; e^{iq\ell},
\label{Cq}
\end{equation}
where $|0\rangle$ is the ground state. 
\begin{figure}
\begin{center}
\epsfxsize=8.5cm
\epsffile{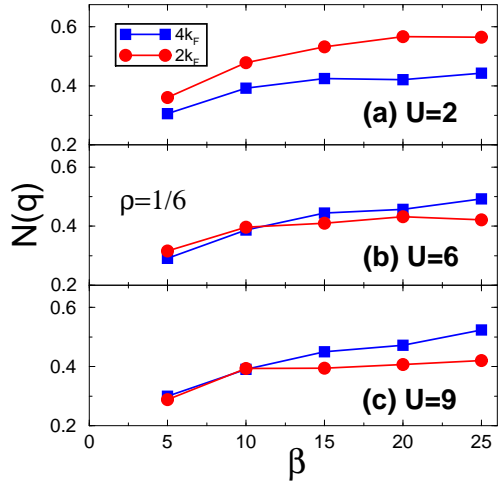}
\caption{QMC results for the charge susceptibility as a function of 
the inverse temperature $\beta$ for a chain with $N_s=36$ sites and
occupation $\rho=1/6$: (a) $U=2$, (b) $U=6$, and (c) $U=9$.
Circles and squares represent data for $q=2k_F$ and $4k_F$, respectively. 
Error bars are smaller than data points, and all lines are guides to 
the eye only. 
}
\label{nbeta16} 
\end{center}
\end{figure}

\begin{figure}
\begin{center}
\leavevmode
\epsfxsize=8.5cm
\epsffile{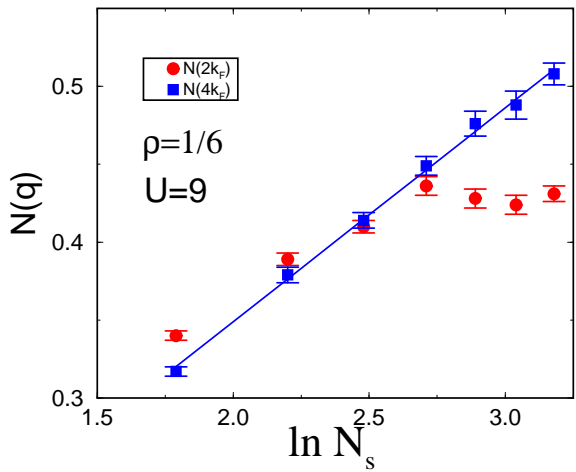}
\caption{QMC results for the charge susceptibility as a function of 
$\ln N_s$ for $U=9$; each point is obtained at $\beta=N_s/4$.
Circles and squares represent data for $q=2k_F$ and $4k_F$, respectively.
The straight line through data for $4k_F$ is a least squares fit.}
\label{nn16}  
\end{center}
\end{figure}

Let us first discuss results for an electronic density $\rho=1/6$,
for which we found that the fermionic determinants in QMC 
simulations do not suffer from the `minus-sign problem';\cite{lg92,vdl92}
this allowed us to reach inverse temperatures as large as $\beta=25$.
>From Fig. \ref{nbeta16} we see that the $4k_F$ charge susceptibility  
appears to be increasing with decreasing temperature, at a 
rate faster than that at $2k_F$, especially for $U=6$ and 9.
The data in Fig.\ \ref{nbeta16} were obtained for $\Delta\tau=0.125$,
but we have explicitly tested other values to ensure they do not  
change significantly as $\Delta\tau\to 0$;
each datum point involves typically 20,000 QMC sweeps 
over all time slices.
Further, in order to check if this increase is limited by finite-size or 
finite-temperature effects, we performed additional simulations on a
`square space-time lattice',\cite{HS} i.e., we set $N_s=L$; 
the result is displayed in Fig.\ \ref{nn16} for $N_s=L\leq 96$. 
While the charge susceptibility at $2k_F$ seems to saturate as 
$N_s$ increases, the one at $4k_F$ still scales with $\ln N_s$, 
up to the largest sizes considered. 
Thus, in spite of the very low temperatures reached, we were still 
unable to find indications of a crossover temperature below which the 
system is dominated by the $2k_F$ instability.
Also, since $N_s\propto 1/T$ for the data in Fig.\ \ref{nn16}, 
the $4k_F$ charge susceptibility grows logarithmically with the temperature in 
this range, similarly to the infinite coupling limit.\cite{HS}
\begin{figure}[t]
\begin{center}
\leavevmode
\epsfxsize=8.5cm
\epsffile{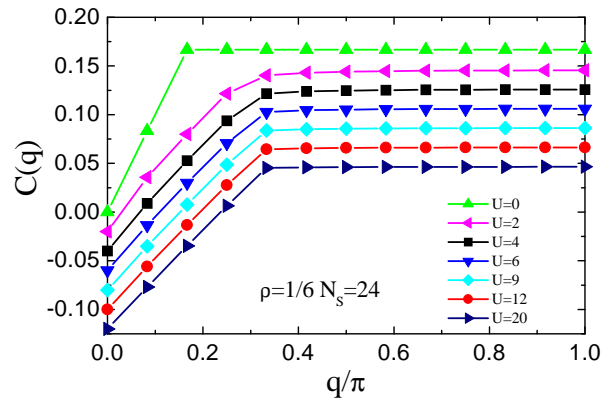}
\caption{Charge structure factor for electron density $\rho=1/6$ and 
$U=0$ (up triangles), 2 (left triangles), 
4 (squares), 6 (down triangles), 9 (diamonds), 12 (circles),
and 20 (right triangles). The system size is $N_s=24$, and 
successive vertical shifts by -0.02 have been imposed on the
curves, for clarity.
}
\label{cq16} 
\end{center}
\end{figure}

\begin{figure}[t]
\begin{center}
\leavevmode
\epsfxsize=8.5cm
\epsffile{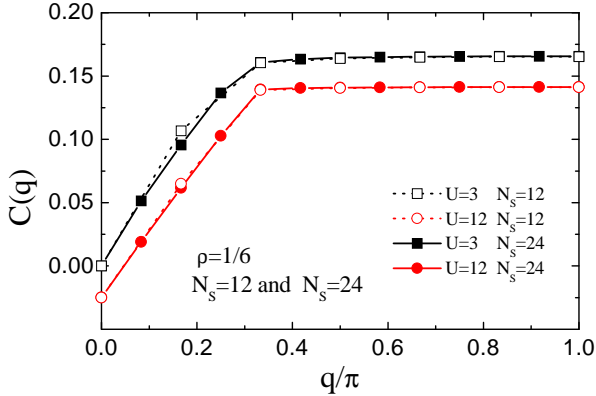}
\caption{Charge structure factor for electron density $\rho=1/6$. 
Empty squares: $U=3,\ N_s=12$;
filled squares: $U=3,\ N_s=24$; 
empty circles: $U=12,\ N_s=12$;
filled circles: $U=12,\ N_s=24$. 
A vertical shift by -0.025 has been imposed on the curves for $U=12$, for clarity.
}
\label{cqfss} 
\end{center}
\end{figure}

We now discuss the charge structure factor at zero temperature, as
obtained from Lanczos diagonalizations, still for $\rho=1/6$. 
Figure \ref{cq16} shows $C(q)$ for several values of $U$;
for clarity, the curves are shown after suffering successive displacements. 
For the free case, $U=0$, we see a sharp plateau beginning at 
$q=2k_F=\pi/6$, which is the signature of the Peierls instability. 
This behaviour is quite different from the one observed for $U \neq 0$:
though somewhat rounded for $U=2$ and 4,
the {\it plateaux} now start at $q=4k_F=\pi/3$.
It is instructive to examine how these roundings evolve
with system size. 
In Fig.\ \ref{cqfss}, we single out data for $C(q)$ with
$U=3$ and 12, and sizes $N_s=12$ and 24.
For each value of $U$, the data below and above $4k_F$ 
respectively move down and up  as $N_s$ increases, thus sharpening the cusp; 
in addition, the position of the latter shows no tendency of shifting 
from $4k_F$.
Thus, the Lanczos results are consistent with those from 
QMC simulations, in the sense that a $4k_F$ instability
is dominant, already for moderate values of $U$.
At this point, it is worth pointing out that the charge structure factor
for the Hubbard chain with second neighbour hopping has been calculated through
density matrix renormalization group (DMRG).\cite{Daul98}
Though their interest was to extract $K_\rho$ from the slope of $C(q)$
at $q=0,$
if one reinterprets those data along the lines discussed here, a
predominance of the $4k_F$ instability can be clearly inferred for the
largest values of $U$ shown in Fig.\ 11 of Ref.\ \onlinecite{Daul98}.
DMRG calculations for the two-leg Hubbard ladder have also led
to $4k_F$-like charge correlations.\cite{Noack96}

\begin{figure}
\begin{center}
\leavevmode
\epsfxsize=8.5cm
\epsfbox{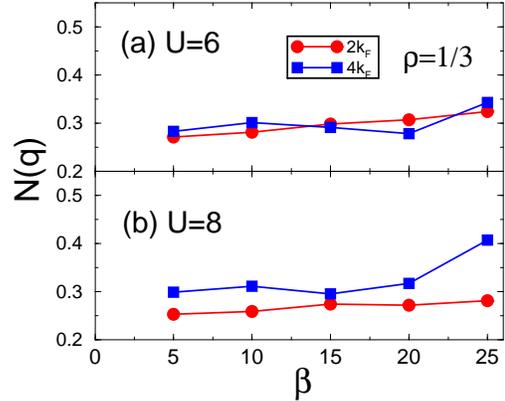}
\caption{QMC results for the charge susceptibility as a function of 
the inverse temperature $\beta$ for a chain with $N_s=36$ sites and 
occupation $\rho=1/3$:  (a) $U=6$ and (b) $U=8$.
Circles and squares represent data for $q=2k_F$ and $4k_F$, respectively.
Error bars are smaller than data points, and all lines are guides to 
the eye only.}
\label{nbeta13} 
\end{center}
\end{figure}

\begin{figure}
\begin{center}
\leavevmode
\epsfxsize=8.5cm
\epsfbox{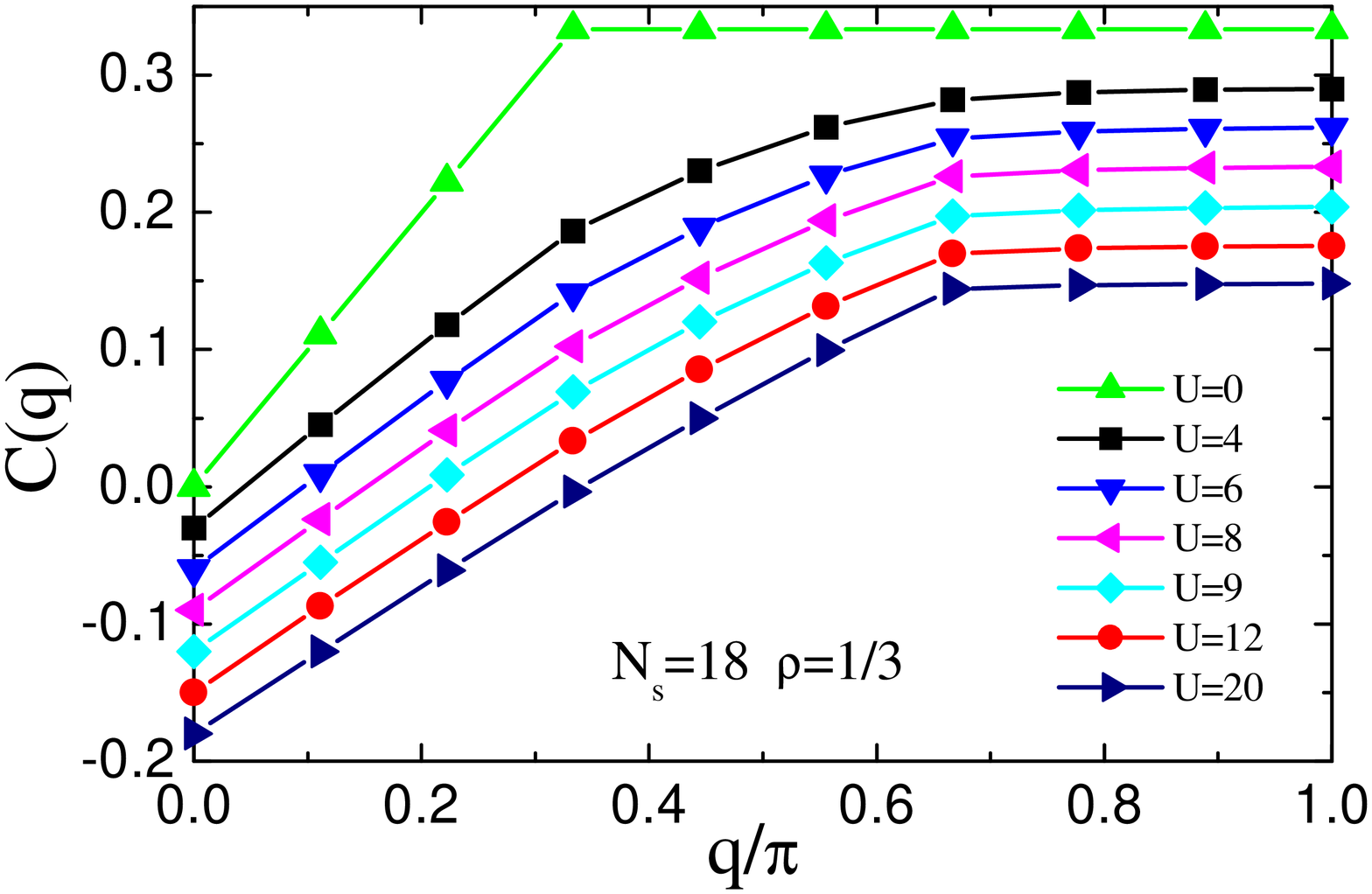}
\caption{Charge structure factor for electron density $\rho=1/3$ and 
$U=0$ (up triangles), 4 (squares), 6 (down triangles), 8 (left triangles),
9 (diamonds), 12 (circles),
and 20 (right triangles). The system size is $N_s=18$, and 
successive vertical shifts by -0.03 have been imposed on the
curves, for clarity.}
\label{cq13} 
\end{center}
\end{figure}

We now change the band filling to $\rho=1/3$; the average sign of the
fermionic determinant is $\sim0.9$ in the worst cases, thus posing no problems
to the resulting averages. 
Figure \ref{nbeta13} shows QMC data for the charge susceptibilities
with $U=6$ and $U=8$.
For $U=6$ an upturn at lower temperatures seems to be setting in
for $4k_F$, while the $2k_F$ data show no noticeable change in growth rate.
On the other hand, for $U=8$ the $4k_F$ susceptibility grows unequivocally faster
with $\beta$ than the one at $2k_F$; see Fig.\ \ref{nbeta13}. 
The corresponding Lanczos data for the charge structure factor are shown in Fig.\ \ref{cq13}
and, similarly to Fig.\ \ref{cq16}, 
the cusp at $q=4k_F=2\pi/3$ gets visibly sharper as $U$ increases. 

\begin{figure}[tbp]
\begin{center}
\leavevmode
\epsfxsize=8.5cm
\epsfbox{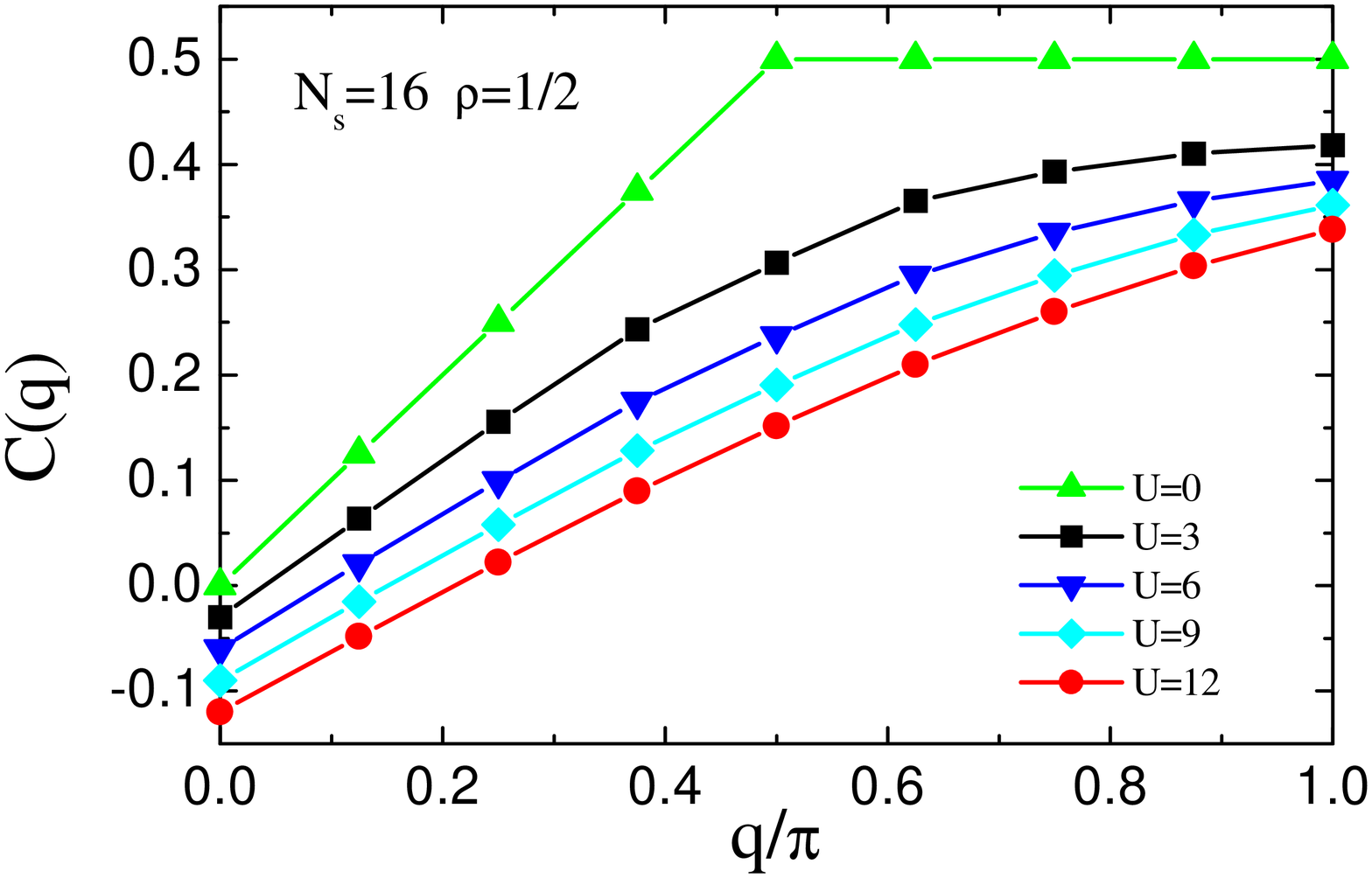}
\caption{Charge structure factor for electron density $\rho=1/2$ and 
$U=0$ (up triangles), 3 (squares), 6 (down triangles), 
9 (diamonds), 12 (circles). 
The system size is $N_s=16$, and 
successive vertical shifts by -0.03 have been imposed on the
curves, for clarity.
}
\label{cq12} 
\end{center}
\end{figure}

Unfortunately, for larger band fillings the `minus-sign problem' prevents us from 
reaching very low temperatures.  
Nonetheless, down to the lowest temperatures probed with acceptable 
average signs of the fermionic determinant 
(i.e., $\langle {\rm sign}\rangle\sim0.7$ at 
$T\sim 1/20$), no indications of a $2k_F$ peak dominating 
the charge susceptibility were found for $\rho=1/2$ or 3/4.
Accordingly, the charge structure factor at $T=0$, 
calculated through Lanczos diagonalizations on a 16-site chain
at quarter filling, shown in Fig.\ \ref{cq12}, confirms the
previous patterns: there is some rounding near 
$q=4k_F=\pi$, which sharpens as $U$ increases, 
consistently with a $4k_F$ instability setting in for finite $U$'s.

In summary, for all band fillings examined, the charge instability seems
to be characterized by a $4k_F$ modulation, rather than by 
$2k_F$, at least for $U$ greater than some $U^*(\rho)$.
The question of how can these findings be reconciled with the analyses of
the continuum model still remains.
Since Lanczos data have been obtained at zero temperature, and
QMC simulations reached much lower temperatures than before,\cite{HS} 
the scenario of a temperature-driven crossover seems now unlikely;
it should be recalled that this crossover was predicted 
based on a {\em weak coupling} RG analysis.
We therefore envisage the following scenario.
While analyses of the Luttinger liquid have so far provided detailed 
insight into the behaviour of the exponent $K_\rho$, little is known
about the dependence of the amplitude $A_1$ of the $2k_F$ contribution
[see Eq.\ (\ref{nn})] with $\rho$ and with the coupling constant. 
Our results may be indicating that, for fixed density $\rho$,
$A_1\to 0$ very fast with increasing $U$, either exponentially or, 
less likely, as 
$\left(U^*(\rho)-U\right)^\psi,$ for $U\leq U^*(\rho)$, with $\psi>1$;
a crude examination of the roundings near $4k_F$ in the 
structure factors is consistent with $U^*$ growing with $\rho$.
An alternative scenario could be that $2k_F$ charge 
correlations in the lattice model would suffer from 
unusually slow finite-size effects, thus hindering any present day
numerical calculations to detect their predominance over the $4k_F$
ones; however, since one would need effects slower than those suggested 
by the logarithmic `correction' in Eq.\ (\ref{nn}), this scenario
is less appealing.
Therefore, the presence of a $4k_F$ instability can be made compatible 
with Luttinger liquid picture through the behaviour of the amplitude
of the $2k_F$ contribution. 
We hope our results stimulate more extensive work both on the
Luttinger liquid and on the  
lattice model in order to extract a quantitative behaviour for
the amplitude $A_1(\rho,U)$. 

\acknowledgments
The authors are grateful to H.\ Ghosh, A.\ L.\ Malvezzi and 
D.\ J.\ Scalapino for useful discussions, 
to S.\ L.\ A.\ de Queiroz and R.\ T.\ Scalettar for
comments on the manuscript, and to R.\ Bechara Muniz for computational 
assistance.
Financial support from the Brazilian Agencies FAPERJ, FINEP, CNPq 
and CAPES 
is also gratefully acknowledged.
The authors are also grateful to Centro de Supercomputa\c c\~ao
da Universidade Federal do Rio Grande do Sul for the use of
the Cray T94, and to the Instituto de F\'\i sica at Universidade
Federal Fluminense, where this work initiated.

\end{document}